\documentstyle[twocolumn,aps,epsfig,here]{revtex}

\begin{document}
\title{
Event-by-event fluctuations of the kaon to pion ratio
in central Pb+Pb  \\ collisions at 158~GeV per Nucleon
}
\author{
S.V.~Afanasiev$^{11}$, T.~Anticic$^{22}$, J.~B\"{a}chler$^{7,9}$, D.~Barna$^{5}$,
L.S.~Barnby$^{3}$, J.~Bartke$^{8}$, R.A.~Barton$^{3}$,
L.~Betev$^{15}$, H.~Bia{\l}\-kowska$^{19}$, A.~Billmeier$^{12}$
% ...
\\
{\bf NA49 Collaboration}}
\address{
% ...
$^{3}$Birmingham University, Birmingham, England.\\
% ...
$^{5}$KFKI Research Institute for Particle and Nuclear Physics, Budapest, Hungary.\\
% ...
$^{7}$CERN, Geneva, Switzerland.\\
$^{8}$Institute of Nuclear Physics, Cracow, Poland.\\
$^{9}$Gesellschaft f\"{u}r Schwerionenforschung (GSI), Darmstadt, Germany.\\
% ...
$^{11}$Joint Institute for Nuclear Research, Dubna, Russia.\\
$^{12}$Fachbereich Physik der Universit\"{a}t, Frankfurt, Germany.\\
% ...
$^{15}$University of California at Los Angeles, Los Angeles, CA, USA.\\
% ...
$^{19}$Institute for Nuclear Studies, Warsaw, Poland.\\
% ...
$^{22}$Rudjer Boskovic Institute, Zagreb, Croatia.\\
% ...
}


\maketitle
%
\noindent
\begin{abstract}
We present the first measurement of fluctuations from event to event in the
production of strange particles in collisions of heavy nuclei. The ratio
of charged kaons to charged pions is determined for individual
central Pb+Pb collisions.
After accounting for the fluctuations due to detector resolution and finite number
statistics we derive an upper limit on genuine non-statistical fluctuations, perhaps
related to a first or second order QCD phase transition. Such fluctuations are shown
to be very small.
%
%
\end{abstract}


Quantum Chromodynamics predicts that at sufficiently high energy density strongly
interacting matter will undergo a phase transition from hadronic matter to
a deconfined state of quarks and gluons, the quark gluon plasma (QGP) \cite{lattice}.
To create and study this state of matter in the laboratory collisions of heavy ions
are studied at the CERN SPS which provides lead ($^{208}$Pb) beams of 158~GeV per nucleon.
Recent data suggest that conditions consistent with the creation of a QCD phase transition
are indeed reached in central Pb+Pb collisions \cite{stockqm99}. Whereas the position of the
phase transformation in temperature, energy density and baryon density may thus be located
we are lacking information as to the nature and order of that transition. These might be reflected
by the presence or absence of fluctuations that are characteristic for a first or second order
phase transition. An early theoretical investigation by Kapusta and Mekjian \cite{strange_hic}
suggested such fluctuations in the kaon-to-pion total-yield ratio, due to supercooling-reheating
fluctuations produced by a predicted large enthalpy difference in the two
phases.The $K/\pi$~ratio was shown to fluctuate by about 10\%,
over the domain of conceivable hadronization temperatures, i.e.
140 $<$T$<$200~MeV \cite{becc}.
This prediction would be experimentally testable if the $K/\pi$~ratio or related
quantities could be quantified for individual
central collision events \cite{sto95}.

In a single central Pb+Pb collision at 158~GeV per nucleon about 2400 hadrons
are created \cite{ferenc}, permitting a statistically significant determination
of momentum space distributions and particle ratios on an event-by-event basis.
Using the NA49 large acceptance hadron spectrometer, which detects about $70\%$ of all charged 
particles, we are able to study event-by-event fluctuations of hadronic observables.
In a previous publication \cite{phipt} we have presented results concerning the
fluctuations of the eventwise mean transverse momentum.
We showed that  these fluctuations are very small in central Pb+Pb collisions, 
and that the data are consistent with
a hadronic gas in thermal equilibrium \cite{stephanov99}.

In this letter fluctuations in the eventwise ratio of the number of charged kaons to the 
number of charged pions ($[K^{+}+K^{-}]/[\pi^{+}+\pi^{-}]$) are investigated, yielding information
on fluctuations in hadrochemical composition \cite{stock_part_had} and on the
strangeness-to-entropy ratio \cite{strange_hic}.
We shall show that, similar to the $\left<p_{T}\right>$~study, the event-by-event
fluctuations of the $K/\pi$~ratio exhibit no significant large-scale fluctuation signal.
We discuss fluctuations predicted in a thermodynamical resonance-gas model, which was
previously used in a study of transverse momentum fluctuations \cite{stephanov99}, and
those expected in various microscopic descriptions of the collision evolution.

A detailed description of the NA49 experiment can be found in \cite{nimpaper}.
We used a data set of central Pb+Pb collisions that were
selected by a trigger on the energy deposited in
the NA49 forward calorimeter. The trigger accepted only the 5\% most
central events, corresponding to an impact parameter range $b < 3.5$~fm.
The event vertex was reconstructed using information from beam position
detectors and a fit to the measured particle trajectories. Only events
uniquely reconstructed at the known target position were used.
In this analysis particles were selected that had at least 30 measured points in
one of the two  Main Time Projection Chambers (MTPC) outside
the magnectic field.

A cut on the extrapolated impact parameter of a particle track at the primary vertex
was used to reduce the contribution
of non-vertex particles originating from weak decays and secondary
interactions. We estimate that about 60~\% of such particles are
rejected by the vertex cuts.
No further acceptance cuts were
made, thus maximising the statistical significance of the event-by-event
particle ratio measurement.

%
The particle identification (PID) in this analysis was based on the measured specific energy
loss ($dE/dx$) in the MTPC using a truncated mean algorithm. Details of the
$dE/dx$ measurement can be found in \cite{nimpaper} and in particular in
\cite{rolandc99}.  As the event-by-event particle identification depends
crucially on the stability of the $dE/dx$ measurement with respect to time,
event multiplicity and possible backgrounds, only the energy loss of the track
in the MTPC was used in this analysis, making use of correction algorithms
developed for this detector \cite{rolandc99}. In particular we observed
a significant multiplicity dependence due to the high charge load
on the TPC readout chambers in central Pb+Pb events.
The procedure we employed to correct these effects
are based on detailed measurements of the electronics response using
laser tracks. They perform a channel-by-channel iterative correction of the
raw TPC charge measurements taking into account the charge history of
sets of neighbouring channels, which are coupled via crosstalk effects
through the sense wires of each TPC readout chamber. These corrections
improved the average $dE/dx$ resolution by about 30\% from
$\sigma_{dE/dx/}\langle dE/dx \rangle = \mbox{5-6}\%$ to the final value of
$3.9\%$ for central Pb+Pb collisions. More importantly, the multiplicity
dependence of the $dE/dx$ measurement was reduced by more than 90\%, leaving
a change of less than $0.3\%$ in the multiplicity range used in this analysis.
\begin{figure}[htp]
\centerline{
\epsfig{file=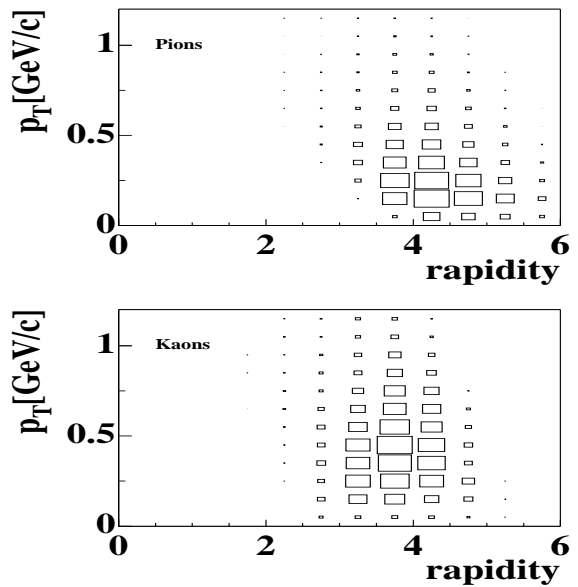,height=8cm,width=8cm}
}
\caption{Phase-space distribution of accepted charged particles pions (above)
and kaons (lower plot). No acceptance correction was applied to the kaon to
pion ratios reported here.}
\label{1}
\end{figure}
Within our acceptance window for kaons and pions (shown in Fig. 1) the $dE/dx$
resolution translates into an average separation of pions from kaons of about 2.1 sigma and
of kaons from protons of about 1.8 sigma.
Obviously, given the accepted multiplicity of about 40 kaons per event, it becomes crucial
to make maximum use of the available information when constructing an estimator
for the event-by-event kaon-to-pion ratio. In particular a simple counting of
particles is not possible under these conditions.
In our analysis we use a PID method closely related to that proposed
in \cite{gazdzicki95}.
 
Combining the information from all events, the momentum distributions normalized
to unity, $F_m(p)$, were determined for each particle species
($m = $ kaons, pions, protons, electrons). 
We also evaluated the normalized probability density
functions for the truncated mean energy loss, 
$f_m (\vec{p};~dE/dx)$, as a function
of particle momentum for each species. 
The relative yield of different particle species
is characterized by parameters $\Theta_{m}$, such that $\sum_m \Theta_{m} = 1$.
These parameters, with the ratio $\Theta_K/\Theta_{\pi}$ giving the
$K/\pi$ ratio of interest here, were determined for every single event by maximizing
the likelihood function
\begin{equation}
L = \prod_{i=1}^n [\sum_m \Theta_{m} F_{m} (p_i) f_{m} (p_i;
(dE/dx)_i)]
\end{equation}
using directly the observed momentum $p_i$, and specific energy loss $(dE/dx)_i$, for each
 particle $i$ in the event. 
The use of fixed momentum distributions $F_m(p)$, implying small changes in
the particle momentum distributions from event to event, is justified by the result
of our analysis of event-by-event fluctuations in transverse momentum
\cite{phipt} which were found to be in the range of 1\% or smaller.
The resulting distribution of event-by-event $K/\pi$ ratios is shown in
Fig.~2 (points). The shape of the distribution can be understood as the
result of three main contributions:

Firstly, due to the finite number of particles produced and observed per event,
the ratio of particle multiplicities measured event-by-event will exhibit
statistical fluctuations with a width dictated by the individual particle
multiplicities.

Secondly, due to non-ideal particle identification these pure number 
fluctuations
will be smeared by the experimental $dE/dx$ resolution and the event-by-event
fitting procedure outlined above.

Finally, superimposed on the statistical and experimental fluctuations we expect to observe
any true non-statistical fluctuations. The characterization of the strength of these
non-statistical fluctuations is the goal of this analysis.

In order to isolate the strength of non-statistical fluctuations in the event-by-event
$K/\pi$ distribution, we need to establish a reference providing us with the size
of the contributions from finite number statistics and experimental resolution,
but not containing any further correlations between particles.
This was achieved using a mixed-event reference sample. The mixed
events were constructed by combining particles randomly selected from different
real events, reproducing the multiplicity distribution of the real events.
Only one track of each event was used in the mixing to exclude any residual
correlations. Within our central event sample no further selection on
impact parameter was made.
By construction, the mixed events on average have the same kaon to pion
ratio as the real events, but no internal correlations. 
Due to the constraint on the overall
multiplicity distribution they give an accurate estimate of finite-number
fluctuations in the kaon and pion multiplicities. 
Equally importantly,
each track in the mixed events represents an actual $dE/dx$ measurement.
\begin{figure}[htp]
\centerline{\epsfig{file=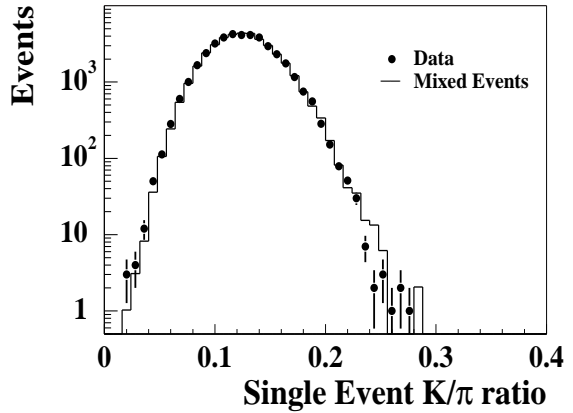,height=8cm,width=6cm,angle=-90}}
\caption{ Distribution of the event-by-event kaon to pion ratio estimated
using a maximum likelihood method (points). As a reference, the same
procedure was applied to a mixed event sample (histogram).}
\label{2}
\end{figure}
The mixed events therefore automatically include all the effects 
of detector resolution.
They are subjected to the same maximum likelihood
fit procedure as the real events, allowing a direct comparison with the data.
Any deviation of the distribution for the data from the purely
statistical mixed-event distribution would indicate the presence of
non-statistical fluctuations in the production of kaons and pions.
The result of this comparison is shown in Fig. 2, showing the distribution
of the $K/\pi$ ratio
for data (points) and for mixed events (histogram).
The small difference between data and statistical reference
immediately illustrates that any correlations or
anti-correlations in the final multiplicities are small. In
particular no significant number of events with unusually
small or large strangeness content is present in our data sample.

The observed relative width of the data distribution is $\sigma_{data} = 23.27\%$.
The width of the mixed event distribution is $\sigma_{mixed} = 23.1\%$. A more
detailed analysis of the latter \cite{rolandc99} shows that it is composed
of contributions from finite number statistics of $\sigma = 15.9\%$
and the $dE/dx$~resolution plus fitting procedure of $\sigma = 16.7\%$
which add in quadrature to the observed width.
Note that the average eventwise $K/\pi$~ratio shown in Fig. 2 does not
necessarily correspond to the true ratio in $4\pi$ acceptance \cite{ferenc},
since no corrections for efficiency and $y,p_{T}$ acceptance were included in the
analysis presented.

To quantify the remaining difference between data and mixed events
we define the strength of non-statistical fluctuations as
\begin{equation}
\sigma_{non-stat} = \sqrt{\sigma_{data}^2 - \sigma_{mixed}^2}
\end{equation}
In general, processes leading to a correlated production of one or the other particle
species or to a correlation in their multiplicities would
result in $\sigma_{non-stat} > 0$.

For our data set we obtain
\begin{equation}
\sigma_{non-stat} = 2.8\% \pm 0.5 \%.
\end{equation}
The fluctuations observed in the data are very small compared to
the scale given by the observed two-fold enhancement of strangeness in nucleus-nucleus
collisions relative to nucleon-nucleon collisions \cite{ferenc}. To a very good
accuracy we can conclude that the mechanism responsible for the enhancement of
strangeness is therefore active in each central Pb+Pb collision.
%

The observed value of $\sigma_{non-stat}$ also allows us to establish a limit on
event-by-event fluctuations as a function of the frequency of occurrence of these fluctuations.
In establishing this limit at the 90\% confidence level we took into account
our estimated systematical uncertainties in determining the event-by-event
$K/\pi$ ratio. As mentioned previously, all systematic uncertainties considered
tend to increase the observed value.
According to our measurement fluctuations of a relative amplitude of $\sigma_{non-stat} > 4.0\%$
can be ruled out in case all events of the sample exhibit the fluctuation pattern, whereas
fluctuations occurring in $5\%$ of the events can only be ruled out for $\sigma_{non-stat} > 15.0\%$
(still small compared to the strangeness enhancement).
 
Finally, we would like to compare our measurement to the
FRITIOF model of nucleus-nucleus collisions. This microscopic
model quantitatively reproduces the correlations observed
in nucleon-nucleon collisions and predicts a value
of $\sigma_{non-stat}^{FRITIOF} = 9\%$ for central Pb+Pb collisions. 
Clearly the fluctuations
in the data are much smaller. The disappearance of fluctuations
when going to central nucleus-nucleus collisions can be
interpreted as evidence for statistical particle production,
as opposed to the highly correlated processes observed in
nucleon-nucleon collisions.
 
This observation lends further support to the interpretation
of particle ratios in nucleus-nucleus collisions using
statistical hadronization models \cite{stockqm99}, with the added information
that the statistical distribution is actually realized in
every individual event and not just on an ensemble basis.

The observed fluctuations can also be compared to the results
of an equilibrium hadron gas calculation \cite{koch99} which includes
correlations induced by many-body decays of resonances
into pions and kaons. This calculation predicts non-statistical
fluctuations $\sigma_{non-stat} \approx 2\%$, in good agreement
with the data.


In conclusion we have presented the first event-by-event measurement of
particle ratios in ultra-relativistic heavy ion collisions.
Using particle identification by $dE/dx$ in the NA49 TPCs
the fluctuations in strangeness production were studied
using the ratio of charged kaons to pions.
No non-statistical fluctuations are observed and we deduce
an upper limit of $\sigma_{non-stat} < 4.0\%$ for fluctuations
occurring in every event at the $3\sigma$-level. The fluctuations are
therefore very small relative to the two-fold strangeness enhancement,
indicating that the dynamical evolution of individual events proceeds
in a very similar fashion. As in our previous study of transverse
momentum fluctuations in Pb+Pb collisions \cite{phipt}, we find no
evidence of fluctuations that might indicate a strong first-order
phase transition or freeze-out near a proposed QCD
critical end point.

On the contrary, the fluctuations observed in Pb+Pb collisions are
significantly smaller than those expected for an indepedent
superposition of nucleon-nucleon collisions. This supports
the interpretation of flavor ratios in terms of statistical
hadronization models, combined with a smooth transition from
a possible partonic state to the final state hadronic particle
composition. In fact, the minimal fluctuations expected due
to resonance production of the final state hadrons seem to
completely exhaust the observed strength of non-statistical fluctuations.

This work was supported by the Director, Office of Energy Research, 
Division of Nuclear Physics of the Office of High Energy and Nuclear Physics 
of the US Department of Energy (DE-ACO3-76SFOOO98 and DE-FG02-91ER40609), 
the US National Science Foundation, 
the Bundesministerium f\"ur Bildung und Forschung, Germany, 
the Alexander von Humboldt Foundation, 
the UK Engineering and Physical Sciences Research Council, 
the Polish State Committee for Scientific Research (2 P03B 02615, 01716, and 09916), 
the Hungarian Scientific Research Foundation (T14920 and T23790),
the EC Marie Curie Foundation,
and the Polish-German Foundation.


\begin{thebibliography}{99}
\bibitem{lattice} See e.g.\ E. Laehrmann,  Nucl. Phys. {\bf A610} (1996) 1; 
F.\ Karsch, Nucl. Phys. {\bf A590} (1995) 367.
\bibitem{stockqm99} R.\ Stock, Nucl. Phys. {\bf A661} (1999) 282c. 
L.\ Kluberg, ibidem p. 300c.
\bibitem{strange_hic}J.\ I.\ Kapusta, A.\ Mekjian, Univ. of Minnesota Supercomputer
Inst. preprint 85/8 (1985), and Phys. Rev. {\bf D33} (1986) 1304.
\bibitem{becc} F. Becattini, private communication.
\bibitem{sto95} R. Stock,  Proceedings of a NATO Advanced Research Workshop
on {\it Hot Hadronic Matter: Theory and Experiment}, 1994, Divonne, France.
\bibitem{ferenc} F.\ Sikler,  Nucl. Phys. {\bf A661} (1999) 45c.
\bibitem{phipt} H.\ Appelsh\"auser et al., Phys. Lett. B {\bf 459} (1999) 679.
\bibitem{stephanov99} M.\ Stephanov, K.\ Rajagopal, E.\ Shuryak,  Phys. Rev. {\bf D60} (1999) 114028.
\bibitem{stock_part_had} R.\ Stock, Phys. Lett. {\bf B456} (1999) 277.
\bibitem{nimpaper} S.\ Afanasiev et al. (NA49 collab.), Nucl. Instr. and Meth. in Phys. Res. {\bf A430} (1999) 210.
\bibitem{rolandc99} C.\ Roland, PhD Thesis University Frankfurt (1999); 
http://na49info.cern.ch/cgi-bin/wwwd-util/NA49/\\NOTE?219
\bibitem{gazdzicki95} M.\ Gazdzicki, Nucl. Instr. and Meth. in Phys. Res. {\bf A345} (1994) 148.
\bibitem{koch99} S.\ Jeon, V.\ Koch, Phys. Rev. Lett. {\bf 83} (1999) 5435.
%
\end{thebibliography}
\end{document}